\documentclass[aps,prc,reprint,showpacs,groupedaddress,onecolumn]{revtex4}
\usepackage{bm}% bold math \usepackage{amsmath} \usepackage{amsfonts}
\usepackage{amscd} \usepackage{graphicx} \usepackage{color}
\usepackage{graphicx}
\usepackage{amsmath}
\usepackage{amsfonts}
\usepackage{amssymb}
\usepackage{amscd}
\def\beq{\begin{equation}}
\def\eeq{\end{equation}}
\begin{document}

\title{Multi-scale methods in quantum field theory}

\author{W.~N.~Polyzou} \affiliation{Department of Physics and
Astronomy, The University of Iowa, Iowa City, IA 52242}

\author{Tracie Michlin} \affiliation{Department of Applied Mathematics and Computational Sciences, The University of Iowa, Iowa City, IA 52242}

\author{Fatih Bulut} \affiliation{Department of Physics, In\"on\"u University,
Turkey}

\date{\today}

\pacs{}

\begin{abstract}

  Daubechies wavelets are used to make an exact multi-scale
  decomposition of quantum fields.  For reactions that involve a
  finite energy that take place in a finite volume, the number of
  relevant quantum mechanical degrees of freedom is finite.  The
  wavelet decomposition has natural resolution and volume truncations
  that can be used to isolate the relevant degrees of freedom.  The
  application of flow equation methods to construct effective theories
  that decouple coarse and fine scale degrees of freedom is
  examined.

\end{abstract}
\maketitle

\section{Introduction}
\label{intro}

Daubechies wavelets \cite{Daubechies}\cite{Daubechies2} are used to
decompose quantum fields into localized degrees of freedom on all
distance scales.  For reactions that involve a finite energy that take
place in a finite volume, the number of relevant quantum mechanical
degrees of freedom is finite.  While a truncation to these degrees of
freedom leads to a mathematically well-defined framework, a realistic
treatment of the dynamics may still require a prohibitively large
number of degrees of freedom for computation.

The Daubechies wavelet decomposition has natural resolution and volume
truncations that can be used to identify the relevant degrees of
freedom.  It is desirable to construct an effective theory that
eliminates degrees of freedom that are important for a realistic
treatment of the dynamics, but not directly related to the scales of
experimental interest.  The application of flow equation methods to
construct effective theories that decouple degrees of freedom with
different resolutions is examined.

%Your text comes here. Separate text sections with
\section{Daubechies basis}
\label{sec:1}
%Text with citations \cite{RefB} and \cite{RefJ}.
%\subsection{Daubechies basis}
%\label{sec:2}
The basis functions are constructed from the fixed point, $s(x)$, of the 
renormalization group equation
\beq
{s(x)} = \underbrace{D
\underbrace{(\sum_{l=0}^{2K-1} h_l T^l {s(x)})}_{\mbox{block average}}
}_{\mbox{rescale}} 
\label{eq:1}
\eeq
where $T$ and $D$ are unitary translation and dilatation operators 
\[
(Tf)(x) = f(x-1) \qquad (Df)(x) = \sqrt{2} f(2x).
\]
The weights $h_l$ are real numbers determined by the conditions that
the functions $s_n(x):=s(x-n)$ are orthonormal and locally finite
linear combinations of these functions can pointwise represent
polynomials of degree $m<K$.  In this work $K=3$ is chosen.  For this
choice $s(x)$ has one continuous derivative and has compact support on
$[0,5]$.  The condition $\int s(x) dx =1$ is imposed on the fixed
point $s(x)$ to fix an initial scale.
%The dimensionless variable $x=y/a$ can be
%thought of as a displacement in units of size $a$.  
Functions of finer resolution are constructed by applying powers of
$D$ to the $s_n(x)$:
%
%\begin{table}[t]
%{\color{black}  
%\caption{\bf \color{black} Filter weights for Daubechies K=3 Wavelets}
%\begin{tabular}{|l|l|}
%\hline				      		      
%$h_0$ & $(1+\sqrt{10}+\sqrt{5+2\sqrt{10}}\,)/16\sqrt{2}$ \\
%$h_1$ & $(5+\sqrt{10}+3\sqrt{5+2\sqrt{10}}\,)/16\sqrt{2}$ \\
%$h_2$ & $(10-2\sqrt{10}+2\sqrt{5+2\sqrt{10}}\,)/16\sqrt{2}$ \\
%$h_3$ & $ (10-2\sqrt{10}-2\sqrt{5+2\sqrt{10}}\,)/16\sqrt{2} $ \\
%$h_4$ & $(5+\sqrt{10}-3\sqrt{5+2\sqrt{10}}\,)/16\sqrt{2}$ \\
%$h_5$ & $(1+\sqrt{10}-\sqrt{5+2\sqrt{10}}\,)/16\sqrt{2}$ \\
%\hline
%\end{tabular}
%\label{coef}
%}
%\end{table}
\[
s^k_n (x) := D^k T^n s(x) = 2^{k/2} s\left (2^k(x-2^{-k}n)\right ) .
\]
These functions have support on intervals of width $(2K-1) \times 2^{-k}$, 
%=5 \times 2^{-k}$.
They are orthonormal and span a 
resolution $2^{-k}$ linear subspace of 
$L^2(\mathbb{R})$ defined by:
\[
{\cal S}_k := \{ f(x) \vert f(x)= \sum_{n=-\infty}^\infty c_n s^k_n(x),
\quad \sum_{n=-\infty}^\infty \vert c_n \vert^2 < \infty \}.
\]
Equation (\ref{eq:1}) implies that 
%\[
%{\cal S}_k := D^k {\cal S}_0\qquad
%\]
the subspaces for different resolutions are nested
\[
{\cal S}_{k} \subset {\cal S}_{k+n} \qquad n\geq 0 
\]
and successive subspaces have non-trivial orthogonal complements
\[
{\cal S}_{k+1} = {\cal S}_{k} \oplus {\cal W}_{k} \qquad {\cal W}_k \not= \{\emptyset \}.  
\]
Iterating this identity gives the exact decomposition of $L^2(\mathbb{R})$ 
into mutually orthogonal subspaces of different resolutions 
\beq
L^2 (\mathbb{R}) = {\cal S}_{k}\oplus {\cal W}_{k} \oplus {\cal W}_{k+1}
\oplus {\cal W}_{k+2} \oplus {\cal W}_{k+3} \oplus  \cdots  .
\label{eq:2}
\eeq
%
%
%\[
%{\cal S}_{k+1} = {\cal S}_{k} \oplus {\cal W}_{k} 
%\]
% 
%
%\[
%L^2 (\mathbb{R}) = {\cal S}_{k}\oplus {\cal W}_{k} \oplus {\cal W}_{k+1}
%\oplus {\cal W}_{k+2} \oplus {\cal W}_{k+3} \oplus  \cdots  =
%\]
%
%\[
%\cdots \oplus {\cal W}_{k-2}\oplus {\cal W}_{k-1} \oplus {\cal W}_{k}
%\oplus {\cal W}_{k+1} \oplus {\cal W}_{k+2} \oplus  \cdots
%\]
An orthonormal basis for ${\cal W}_{k}$ is given by  
functions $\{ w^k_n (x)\}_n$,  
that are constructed from 
the fixed point $s(x)$ by 
\[
w^k_n (x) := D^k T^n w(x) = 2^{k/2} w\left (2^k(x-2^{-k}n)\right ) 
\]
where 
\[
w(x) :=\sum_{l=0}^{2K-1} g_l T^l s(x)
\qquad \mbox{and} \qquad g_l = (-)^l h_{2K-1-l}.
\]
It follows from (\ref{eq:2}) that for any fixed starting resolution, 
$2^{-k}$, the functions 
\beq
\xi_{\mathbf{n}}(\mathbf{x}):= \xi_{n_1}(x)\xi_{n_2}(y) \xi_{n_3}(z)
\qquad
\xi_n(x) \in \{ s_n^k(x), w^l_n(x) \}_{n=-\infty;l\geq k}^{\infty} 
\label{eq:3}
\eeq
are an orthonormal multi-resolution basis of $L^2(\mathbb{R}^3)$ of 
functions that have compact support and one continuous derivative.

\section{Multiresolution decomposition of quantum fields}
\label{sec:2}

The basis (\ref{eq:3}) can be used to decompose the 
quantum fields by resolution \cite{fatih}\cite{tracie}:
\beq
\Phi (\mathbf{x},t) =
\sum_{\mathbf{n}} \Phi^k (\mathbf{n},t) \xi_{\mathbf{n}}(\mathbf{x})   
\qquad
\Phi^k (\mathbf{n},t) = \int d\mathbf{x}\xi_{\mathbf{n}}(\mathbf{x}) 
\Phi (\mathbf{x},t) .
\label{eq:4}
\eeq
While the basis functions are not Schwartz functions, direct computation 
shows that the resulting discrete fields are well-defined operators in the 
free-field case.  
%In additon, the support condition implies that the 
%Fourier transforms of the basis functions are analytic.  

Given a pair of canonical fields, $ \Phi (\mathbf{x},t)$ and 
$\Pi (\mathbf{x},t)$, that satisfy canonical equal-time 
commutation relations,  the corresponding discrete fields will satisfy 
discrete equal-time canonical commutation relations: 
%
%
%
%\vfill
%\[
%\Pi (\mathbf{x},t) =
%\sum_{\mathbf{n}} \Pi^k (\mathbf{n},t)\xi_{\mathbf{n}}(\mathbf{x})  
%\qquad
%\Pi^k (\mathbf{n},t) = \int d\mathbf{x}\xi_{\mathbf{n}}(\mathbf{x}) 
%\Pi (\mathbf{x},t)
%\]
%\vfill
\[
[\Phi (\mathbf{n},t), \Pi (\mathbf{m},t)]=
i \delta_{\mathbf{n},\mathbf{m}}
\]
with all other commutators vanishing.  
%\vfill
%\[
%[\Phi (\mathbf{n},t), \Phi (\mathbf{m},t)]=
%[\Pi (\mathbf{n},t), \Pi (\mathbf{m},t)]=0 , 
%\]
These expansions can be used in field-theoretic Hamiltonians.  In this
discrete representation the integral over the Hamiltonian density is
replaced by an infinite sum.  Terms in a typical Hamiltonian are
replaced by the infinite sums
\[
\mu^2 \int d\mathbf{x} \Phi^2(\mathbf{x},0) \to 
\mu^2 \sum_{\mathbf{m}}  \Phi^2(\mathbf{m},0)  
\qquad
\int d\mathbf{x} \Pi^2(\mathbf{x},0) \to 
\sum_{\mathbf{m}}  \Pi^2(\mathbf{m},0)  
\]
\[
\int d\mathbf{x} \pmb{\nabla}\Phi(\mathbf{x},0) \cdot
\pmb{\nabla}\Phi(\mathbf{x},0) \to 
\sum_{\mathbf{m}\mathbf{n}} D_{\mathbf{m}\mathbf{n}} 
\Phi (\mathbf{m},0)\Phi (\mathbf{n},0)
\]
\[
\lambda \int d\mathbf{x} \Phi^n(\mathbf{x},0) 
\to \lambda 
\sum_{\mathbf{m}_1 \cdots \mathbf{m}_n} \Gamma_{\mathbf{m}_1 \cdots \mathbf{m}_n} 
\Phi (\mathbf{m}_1,0) \cdots \Phi (\mathbf{n}_n,0)
\]
where
\beq
D_{\mathbf{m}\mathbf{n}}:=
\int d\mathbf{x} \pmb{\nabla} \xi_{\mathbf{m}} (\mathbf{x}) \cdot \pmb{\nabla} 
\xi_{\mathbf{n}} (\mathbf{x}) 
\qquad
\mbox{and}
\qquad 
\Gamma_{\mathbf{n}_1 \cdots \mathbf{n}_N} =
\int d\mathbf{x}
\xi_{\mathbf{n}_1} (\mathbf{x}) \cdots \xi_{\mathbf{n}_N} (\mathbf{x}) 
\label{eq:5}
\eeq
are constant coefficients.  They can be computed exactly using the 
renormalization group equation and properties of the basis 
\cite{beylkin}\cite{fatih} .

\section{Truncated quantum fields}

There are natural volume and/or resolution truncations of quantum
fields represented by a Daubechies expansion.  Truncations are defined by
retaining the terms in the expansion (\ref{eq:4}) whose basis functions
have support in a given volume and is larger than a minimal support volume.
The resulting truncated fields are still differentiable functions of
$x$; they are expressed as a finite sum
\beq
\Phi_T (\mathbf{x},t) =
\sum_{\mathbf{n}\in {\cal I}} \Phi^k (\mathbf{n},t) \xi_{\mathbf{n}}(\mathbf{x}).   
\label{eq:6}
\eeq

A truncated Hamiltonian is defined by replacing the fields in the
exact Hamiltonian by the truncated fields.  This truncation limits the
volume and finest resolution of the theory.  The resulting truncated
Hamiltonian has a finite number of degrees of freedom.  
The scaling properties of the integrals \cite{fatih}, 
\[
D^k_{mn}  =
2^{2k} \sum D^0_{mn}  
\qquad 
\Gamma^k_{n_1 \cdots n_N} =
2^{3k({n\over 2}-1)} \Gamma^0_{n_1 \cdots n_N} 
\]
which follow from (\ref{eq:5}),
lead to an exact renormalization group equation for infinite volume 
truncated Hamiltonians with different resolutions
\[
H^k(\Phi^k, \Pi^k, m^{2k} ,\gamma_N^k) =
2^k H^0(\Phi^0, \Pi^0, 2^{-2k}m^{0} , 2^{k(N-4)}\gamma_N^0) 
\]
where the fields in these two Hamiltonians ($1+1$ dimension) are related by the 
canonical transformation
\[
\Phi^k = \eta \Phi^0 \qquad \Pi^k = \eta^{-1} \Pi^0 \qquad \eta= 2^{-k/2}.  
\]
The vacuum of the truncated theory can be constructed by 
decomposing the canonical fields into creation and annihilation 
parts
\[
a_{\mathbf{n}} := {1 \over \sqrt{2}}(\alpha \Phi_{\mathbf{n}} + 
i {1 \over \alpha} \Pi_{\mathbf{n}})  
\] 
where $\alpha$ is any constant; and then solving the coupled
cluster equations \cite{coester}
\[
e^{-S} H_T e^{S} \vert 0 \rangle_0 =0  \qquad 
\qquad a_{\mathbf{n}} \vert 0 \rangle_0 =0 \qquad 
S= \sum S^m_{\mathbf{n}_1 \cdots \mathbf{n}_n} a^{\dagger}_{\mathbf{n}_1}
\cdots a^{\dagger}_{\mathbf{n}_m}  
\]
for the coefficients $S^m_{n_1 \cdot n_n}$. The vacuum of 
the truncated theory is 
$\vert 0 \rangle  = N e^{S} \vert 0 \rangle_0 $,
where $N$ is a normalization constant.
% and $\vert 0 \rangle_0$
%is defined by
%$a_{\mathbf{n}} \vert 0 \rangle_0=0$. 

The truncated fields are solutions of 
the Heisenberg equations
\[
\dot{\Phi}_{\mathbf{n}}(t) = i [H_T, {\Phi}_{\mathbf{n}}(t)]  
\qquad
\dot{\Pi}_{\mathbf{n}}(t) = i [H_T, {\Pi}_{\mathbf{n}}(t)]   
\]
with initial conditions
\[
[\Phi (\mathbf{n},0), \Pi (\mathbf{m},0)]=
i \delta_{\mathbf{n},\mathbf{m}}
\qquad
[\Phi (\mathbf{n},0), \Phi (\mathbf{m},0)]=
[\Pi (\mathbf{n},0), \Pi (\mathbf{m},0)]=0 . 
\]
The solutions have the form (\ref{eq:4}).
Correlation functions are defined as vacuum expectation values of 
products of the truncated fields.  They are differentiable functions of 
the space-time coordinates.

\section{Effective theories}

A benefit of decomposing fields into degrees of freedom with different
resolutions is that it becomes possible to formulate the problem of
constructing an effective theory with physical-scale degrees of
freedom by eliminating dynamically important small-scale degrees of
freedom.  This can be achieved by a block diagonalization of the
Hamiltonian by resolution.  While defining the exact Hamiltonian
requires renormalization, a truncated Hamiltonian that has all of the
degrees of freedom relevant to a given energy and volume is
well-defined and should accurately describe the system of interest.

The problem of decoupling scales is examined for the case of a 
free-field theory where the scale coupling appears in the constant matrices
$D_{\mathbf{m}\mathbf{n}}$.  The advantage of the free field case is
that a block diagonalization by scale can be performed at the operator
level and nature of the degrees of freedom in each block can be
examined.  This test uses a truncated Hamiltonian for a free scalar
field in 1+1 dimensions.  The fields are truncated to include 16
scaling basis functions, $s_n(x)$, and 16 wavelet basis functions,
$w_n(x)$.  These functions represent degrees of freedom on scales
that differ by factor of 2.

The truncated Hamiltonian has general form 
\[
H_T = {1 \over 2} 
[
(\Pi^s ,\Pi^w) 
\left (
\begin{array}{cc}
I_s & 0\\
0 & I_w 
\end{array}
\right ) 
\left (
\begin{array}{c}
\Pi^s \\
\Pi^w  
\end{array}
\right ) +
\]
\[
(\Phi^s ,\Phi^w) 
\left (
\begin{array}{cc}
\mu^2I_s+D_{s} & {D_{sw}}\\
{D_{ws}}  & \mu^2I_w + D_{w} 
\end{array}
\right ) 
\left (
\begin{array}{c}
\Phi^s \\
\Phi^w  
\end{array}
\right ) 
]
\]
where each block represents a $16\times 16$ matrix and 
the non-diagonal elements $D_{sw}$ and $D_{ws}$ are the
integrals of products of derivatives (\ref{eq:5}) of the basis functions.  The
upper block represents the coarse-scale degrees of freedom while the
lower block represents the fine-scale degrees of freedom.  The goal is
to find a unitarily equivalent Hamiltonian, $H(\lambda)$ 
where the block coupling
terms are absent or small.
 
This is tested \cite{tracie} using flow-equation methods due to Wegner \cite{wegner}
\cite{glazek9}
\[
H(\lambda) = U(\lambda)H(0)U^{\dagger}(\lambda) 
\qquad 
{dU(\lambda) \over d\lambda} = {dU(\lambda) \over d\lambda}
U^{\dagger}(\lambda) U(\lambda) =
K(\lambda) U(\lambda) .
\]
The generator $K(\lambda)$ is chosen to be
\[
K(\lambda) =  {dU(\lambda) \over d\lambda}
U^{\dagger}(\lambda) := [G(\lambda),H(\lambda)]
\]
where $G(\lambda)$ is the block diagonal part of $H(\lambda)$.   This
leads to a differential equation directly for $H(\lambda$) that 
becomes a set of coupled equations for the block diagonal, $G(\lambda)$ and 
coupling parts $H_c(\lambda)$ of $H(\lambda)$:
\[
{dG(\lambda)\over d\lambda} =
[H_c(\lambda),[H_c (\lambda),G(\lambda)]]
\]
\[
{dH_c(\lambda)\over d\lambda} =
-[G(\lambda),[G(\lambda),H_c(\lambda)]] .
\]
These equations can be solved in the bases of eigenstates of $G(\lambda)$ 
and $H_{c}(\lambda)$ respectively 
\beq
G_{mn}(\lambda) = e^{\int_0^\lambda (e_{cm}(\lambda')-e_{cn}(\lambda'))^2 d\lambda'}
G_{mn}(0)
\label{eq:7}
\eeq
\beq
H_{cmn}(\lambda) = e^{-\int_0^\lambda (e_{bm}(\lambda')-e_{bn}(\lambda'))^2 d\lambda'}
H_{cmn}(0).
\label{eq:8}
\eeq
The solutions indicate that the scale-coupling parts of the matrix 
are exponentially suppressed as the flow parameter, $\lambda$, is increased. 

To test this the Hilbert-Schmidt norms of the coupling matrices as
functions of the flow parameter $\lambda$ are computed.  Figures 1 and 2
compare the evolution of the Hilbert-Schmidt norms for a 
representative set of scale
coupling coefficients (Figure 1) to fixed-scale coefficients
(Figure 2).  Figure 1 shows that the scale coupling terms are driven
to 0, but the rate of decrease falls off as the flow parameter is
increased, while Figure 2 shows that Hilbert-Schmidt norm of the 
fixed scale coefficients converges to a finite size. 

Figures 3 and 4 give a more detailed picture of the evolution of each
coefficient, initially (Figure 3) and when $\lambda=20$ (Figure 4).  The first
quadrant shows the $16 \times 16$ matrix of coefficients for the
$\Phi$-$\Phi$ scaling function fields, the next diagonal quadrant shows
the coefficients for the $\Pi$-$\Pi$ scaling function fields.  The third
diagonal quadrant shows the coefficients for the $\Phi$-$\Phi$ wavelet
function fields and the fourth diagonal quadrant shows the
coefficients for the $\Pi$-$\Pi$ scaling function fields.  The 
off-diagonal terms in figure 3 are the coefficients of the scale coupling
terms.  Figure 4 shows that they are driven to 0 for $\lambda=20$.
 
One advantage of the free-field is that the truncated Hamiltonian is
the Hamiltonian for 32 coupled harmonic oscillators.  The block
diagonalization will put oscillators with 16 normal modes into the
coarse-scale block and the other 16 into the fine-scale block.  At the
level of the approximation (there is still a small coupling at
$\lambda=20$) the flow equation method used here to separate scales
put the 16 lowest normal modes in the coarse-scale block and the 16
highest normal modes in the fine-scale block.  This is both the
expected and desired behavior.

While this example demonstrates how flow equation methods can be used
to separate scales, a number of issues remain.  This investigation
showed that increasing the resolution adds higher energy normal modes,
but does not reduce the separation between normal mode frequencies
that one would expect in a continuum limit.  In order to reduce this
separation the truncated volume must be increased.  This means the
continuum limit requires that the resolution and volume limits be
taken together.  Another concern is that as the normal mode frequencies get
closer, the separation of the eigenvalues in (\ref{eq:7}-\ref{eq:8})
will get smaller, resulting much slower convergence of the flow
equation.  A third concern is that for interacting theories each
iteration of the flow equation will generate new many-body
interactions.  In order to control the growth in the number of
generated interactions, the scaling properties of each generated
interaction need to be investigated so irrelevant ones can be
eliminated.  Also, the operator form of the flow equation that worked
for the free field, may not be possible with interactions; however it
can be made to work using projection operators.

The extension to 3-dimensions is straightforward.  Convergence of the 
flow equation method was also established for 0 mass.

\begin{figure}
\begin{minipage}[t]{.45\linewidth}
\centering
\includegraphics[angle=000,scale=.35]{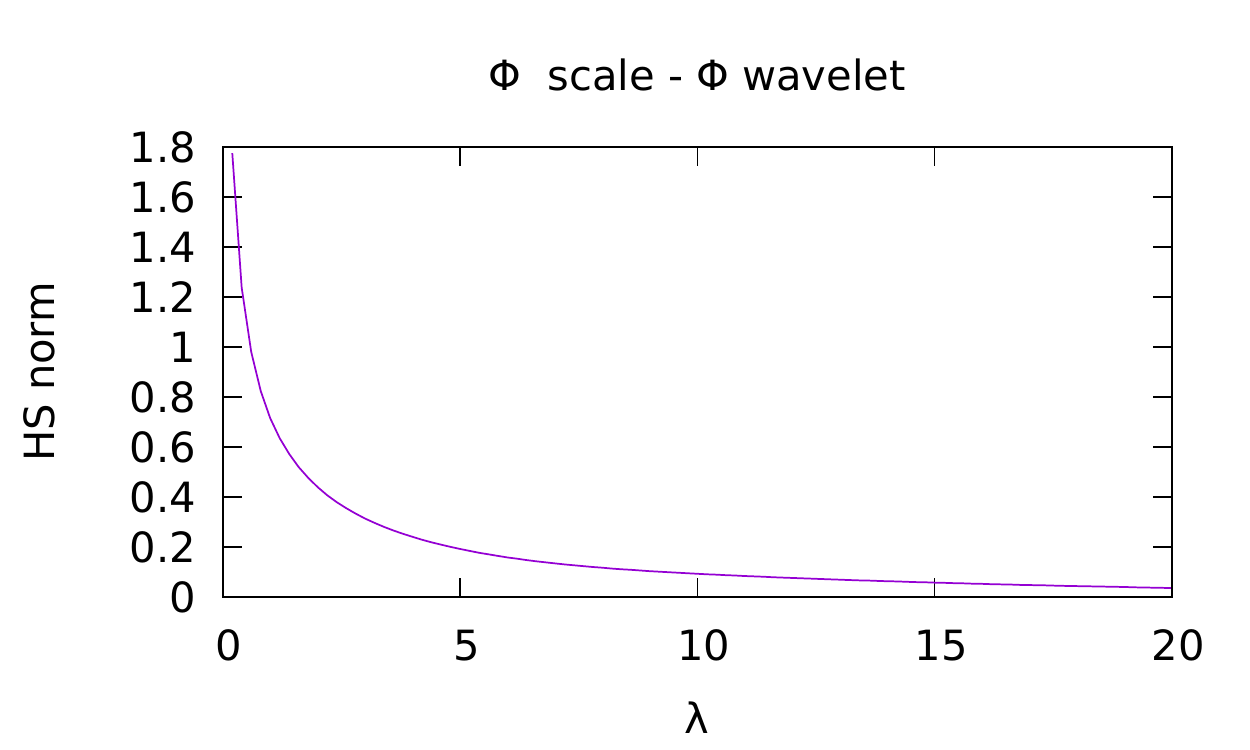}
\caption{$\phi$ scale-$\phi$ wavelet}
\label{fig:3}
\end{minipage}
\begin{minipage}[t]{.45\linewidth}
\centering
\includegraphics[angle=000,scale=.35]{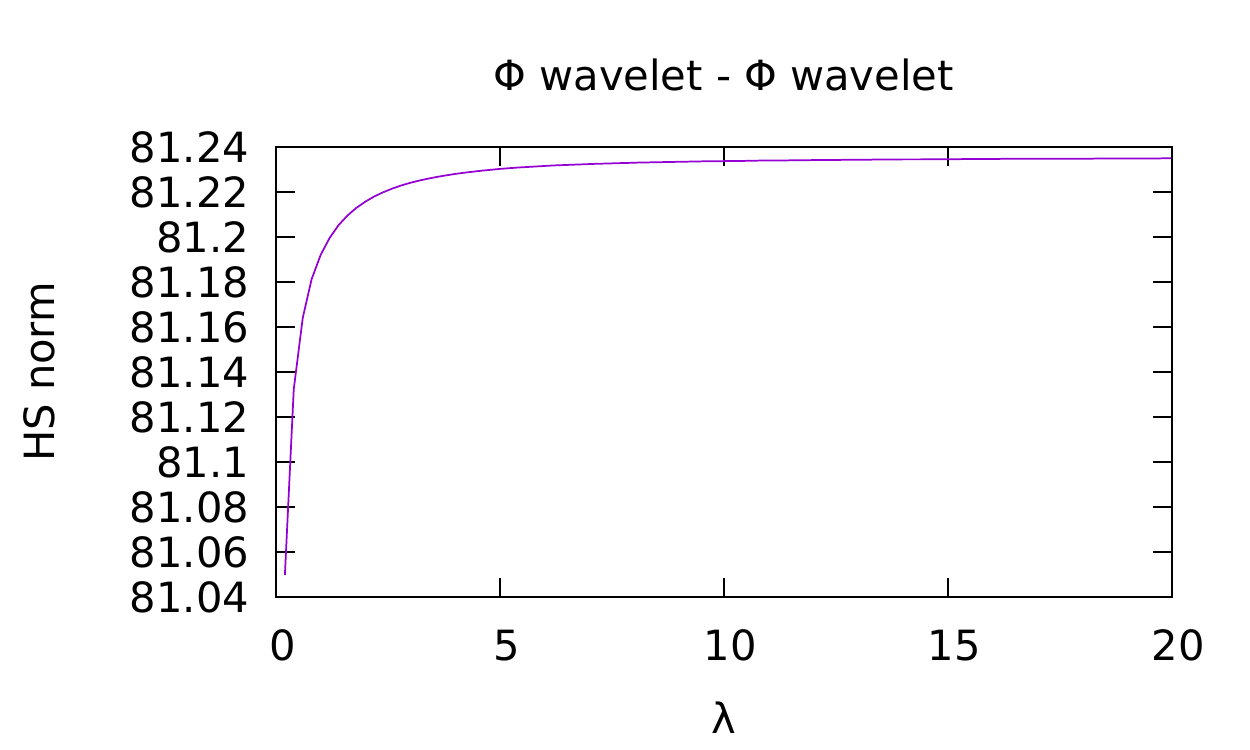}
\caption{$\phi$ wavelet-$\phi$ wavelet}
\label{fig:4}
\end{minipage}
\end{figure}

\begin{figure}
\begin{minipage}[t]{.45\linewidth}
\centering
\includegraphics[angle=000,scale=.25]{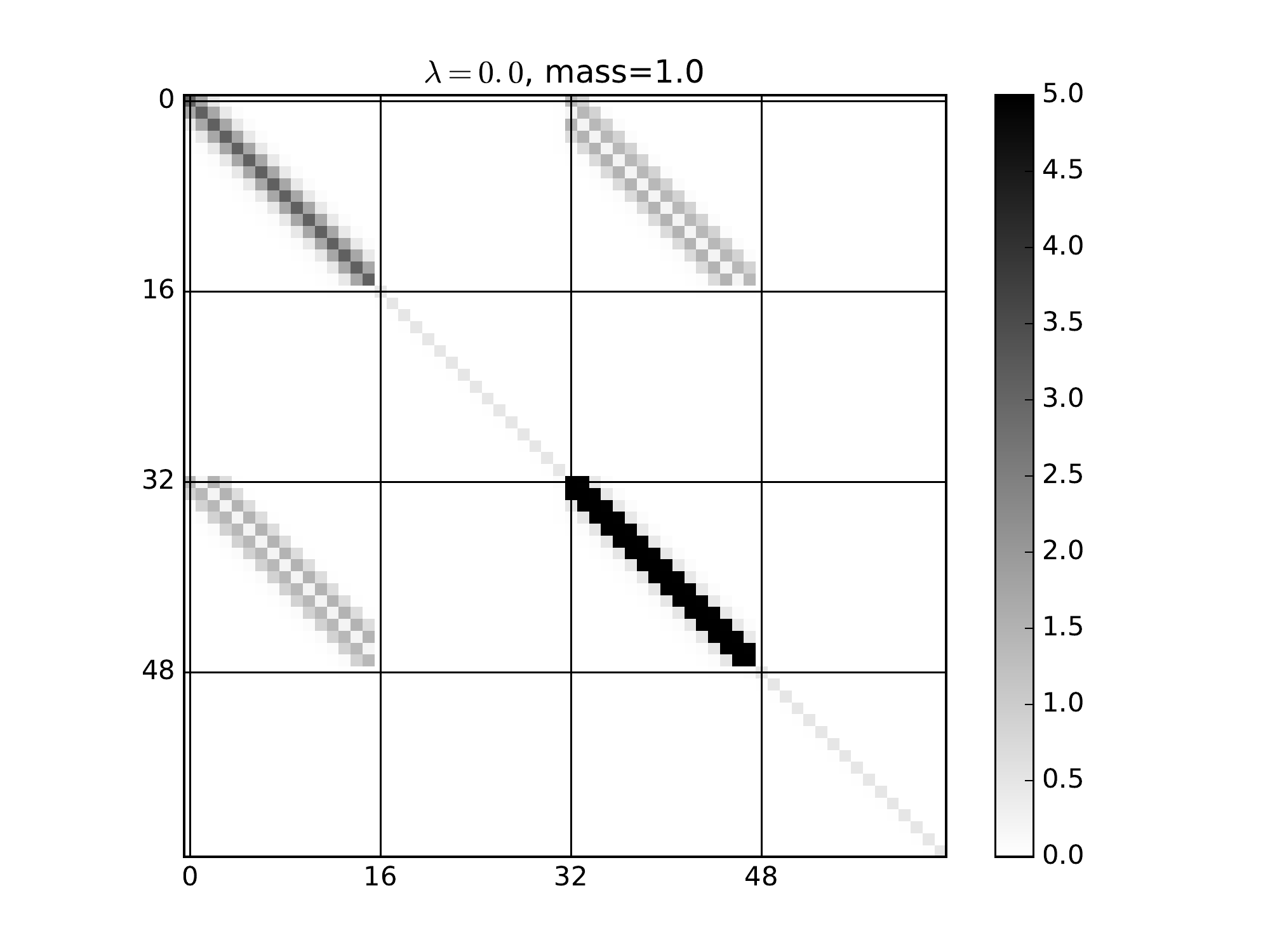}
\caption{Full matrix, $\lambda$=0, mass=1}
\label{fig:1}
\end{minipage}
\begin{minipage}[t]{.45\linewidth}
\centering
\includegraphics[angle=000,scale=.25]{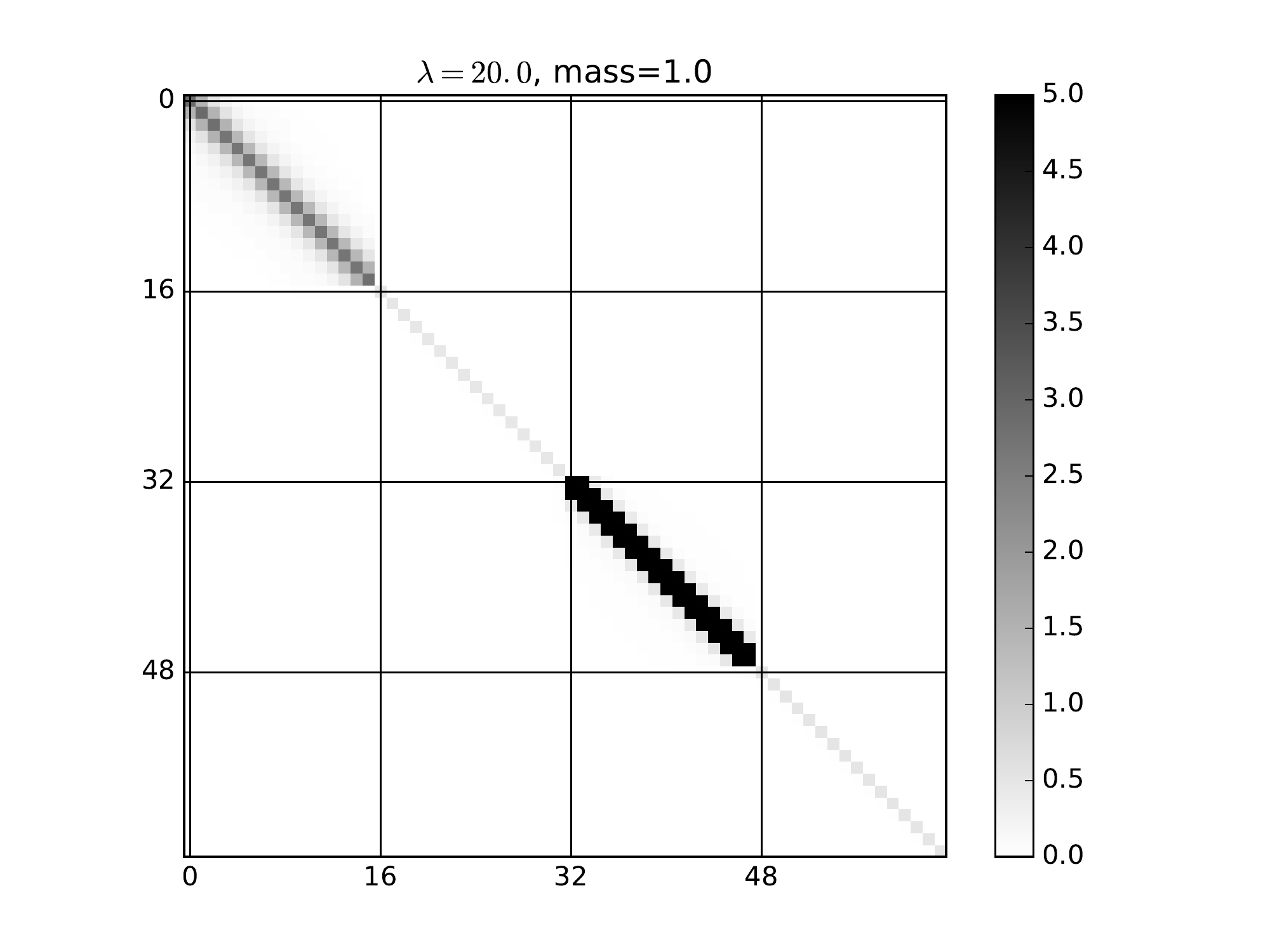}
\caption{Full matrix, $\lambda$=20, mass=1}
\label{fig:2}
\end{minipage}
\end{figure}

W.P. acknowledges the generous support of the U.S. Department of Energy,
grant number DE-SC0016457 who supported this research effort.

\end{document}